\documentstyle[12pt]{article}

\begin{document}
\title{Relations among Elements \\ of the Quark Mass Matrices}
\author{
{K. Harayama}\footnote{E-mail address: harayama@post.kek.jp}\\
{\it Theory Group, KEK, Tsukuba, Ibaraki 305-0801, Japan}
}
\date{}
\maketitle

\begin{abstract}
Current experimental data suggest some relations 
between the Kobayashi-Maskawa Matrix and the quark mass ratios, 
namely 
$|V_{us}| \sim \sqrt{m_d / m_s}$, 
$|V_{ub} / V_{cb}| \sim \sqrt{m_u / m_c}$ and 
$|V_{cb}| \sim m_s / m_b$. 
We consider these relations seriously in this paper. 
As a result, 
they offer us some attractive relations 
among the elements of the mass matrices 
and bring us a problem. 
They might give us informations about the origin of Yukawa couplings. 
\end{abstract}

\section{Introduction}
\par
\hspace*{\parindent}
It is very interesting to consider the quark mass matrices and 
the Kobayashi-Maskawa Matrix (KM) matrix \cite{KM}. 
It is because 
the origin of the quark masses and the KM matrix 
is not understood in the standard model (SM), 
though the SM is able to explain many experimentations. 
The fundamental theory 
is expected to explain the relations among the parameters in the SM. 
For that reason, understanding of relations among the SM parameters 
would offer us important information about the fundamental theory. 

Current experimental data \cite{PDG} give us some hints 
for relations between the KM Matrix and the quark mass ratios, 
namely 
$|V_{us}| \sim \sqrt{m_d / m_s}$, 
$|V_{ub} / V_{cb}| \sim \sqrt{m_u / m_c}$ and 
$|V_{cb}| \sim m_s / m_b$ . 
These relations might be just an accident. 
But they would be important informations about the origin of Yukawa couplings, 
therefore, it is meaningful to stand on such a point of view. 
Starting with a specific theory or symmetry, 
many attempt have been made to derive relations 
between the KM matrix and the quark mass ratios 
using the specific mass matrices with less than 10 parameters 
\cite{Fritzsch1}-\cite{Review}. 
However, 
we start from the KM matrix and the quark masses 
using generic and minimal parametrization of the mass matrices. 
This approach has the advantage of 
considering the origin of quark masses and the flavor mixing. 

Consequently, we can bring 
the relations between elements of the KM matrix and the quark mass ratios 
back to relations among elements of the mass matrices. 
It is not just relations among parameters 
but relations among elements of the mass matrices, 
so it is very attractive 
because they might give suggestions about the origin of Yukawa couplings. 

In addition, 
they give us a problem which can not solve by only flavor symmetry. 
This is common problem when we suppose some natural conditions. 
Under the conditions, we must introduce other mechanism or something. 

This paper is organized as follows. 
In section {\ref{sec:KM}}, 
we make our starting point about the KM matrix clear. 
In section {\ref{sec:Mass_Marix}}, 
we introduce one set of minimally generic quark mass matrices 
and parametrize the elements of them. 
In section {\ref{sec:Discussion}}, 
we consider the relations among the elements of the quark mass matrices. 
Finally, in section {\ref{sec:Summary}}, 
we summarize the results. 

\section{The Kobayashi-Maskawa matrix}
\label{sec:KM}
\par
\hspace*{\parindent}
Current experimental data provide us hints of the relations 
between quark mass ratios and the KM matrix elements \cite{PDG}: 
\begin{eqnarray}
|V_{us}| & \sim & \sqrt{\frac{m_d}{m_s}} ~ \sim ~ 0.22 , 
\nonumber \\
|V_{cb}| & \sim & \frac{m_s}{m_b} ~ \sim ~ 0.03 ~{\mathrm {to}}~ 0.05 , 
\nonumber \\
\left | \frac{V_{ub}}{V_{cb}}\right | & \sim & \sqrt{\frac{m_u}{m_c}} 
 ~ \sim ~ 0.06 ~{\mathrm {to}}~ 0.10 . 
\label{eqn:KM_mass_relations}
\end{eqnarray} 
These relations may be just an accident, 
but we consider the relations seriously. 

The up (down) part mass matrix $M_{\mathrm{u}}$ ($M_{\mathrm{d}}$) 
is diagonalized by unitary matrices $U_{\mathrm{u}}$ and $V_{\mathrm{u}}$ 
($U_{\mathrm{d}}$ and $V_{\mathrm{d}}$), 
\begin{equation}
U_{\mathrm{u}}^{\dagger} M_{\mathrm{u}} V_{\mathrm{u}} = D_{\mathrm{u}} , 
~~~~~
U_{\mathrm{d}}^{\dagger} M_{\mathrm{d}} V_{\mathrm{d}} = D_{\mathrm{d}} , 
\end{equation}
where $D_{\mathrm{u,d}}$ are diagonal matrices. 
Then the KM matrix $V_{\mathrm{KM}}$ is 
\begin{equation}
V_{\mathrm{KM}} = U_{\mathrm{u}}^{\dagger} U_{\mathrm{d}} .
\end{equation}
The KM matrix is almost the unit matrix, 
especially $|V_{ub}|$ and $|V_{td}|$ are very small 
in comparison with the other elements. 
So let us suppose 
that $U_{\mathrm{u}}$ and $U_{\mathrm{d}}$ are almost the unit matrix, 
and that (1,3) and (3,1) elements of $U_{\mathrm{u,d}}$ are very small. 
If we do not suppose them, 
we need large cancellations between up and down part 
to obtain realistic KM matrix. 
But, now, we suppose there is no such cancellation. 
Then the relations (\ref{eqn:KM_mass_relations}) remind us 
that the KM matrix is expressed by quark mass ratios as follows: 
\begin{eqnarray}
V_{\mathrm{KM}} &=& U_{\mathrm{u}}^{\dagger} U_{\mathrm{d}} , 
\nonumber \\ 
U_{\mathrm{u}} &\sim& 
\left( \begin{array}{ccc}
	\exp{\left( \mathrm{i}\theta \right)} & 0 & 0 \\
	0 & 1 & 0 \\
	0 & 0 & 1
\end{array} \right) 
\left( \begin{array}{ccc}
	1 & \displaystyle \sqrt{\frac{m_u}{m_c}} & 0 \\
	\displaystyle -\sqrt{\frac{m_u}{m_c}} & 1 & 0 \\
	0 & 0 & 1
\end{array} \right) , 
\nonumber \\ 
U_{\mathrm{d}} &\sim& 
\left( \begin{array}{ccc}
	1 & 0 & 0 \\
	0 & 1 & \displaystyle \frac{m_s}{m_b} \\
	0 & \displaystyle -\frac{m_s}{m_b} & 1
\end{array} \right) 
\left( \begin{array}{ccc}
	1 & \displaystyle \sqrt{\frac{m_d}{m_s}} & 0 \\
	\displaystyle -\sqrt{\frac{m_d}{m_s}} & 1 & 0 \\
	0 & 0 & 1
\end{array} \right) . 
\label{eqn:KM_mass_ratios}
\end{eqnarray}
On this point of view, 
we consider the quark mass matrices in this paper. 

\section{Quark mass matrices}
\label{sec:Mass_Marix}
\par
\hspace*{\parindent}
If we assume up and down part quark mass matrices are hermitian, 
the expression (\ref{eqn:KM_mass_ratios}) can be constructed 
when the up and down part mass matrices $M_{\mathrm{u}}$ and $M_{\mathrm{d}}$ 
have the following form: 
\begin{eqnarray}
M_{\mathrm{u}} &=& \left( \begin{array}{ccc}
      a_{\mathrm{u}}  & d_{\mathrm{u}}\exp{\left( {\mathrm{i}}\theta \right)} & 0         \\
      d_{\mathrm{u}}\exp{\left( -{\mathrm{i}}\theta \right)} & b_{\mathrm{u}} & 0         \\
      0          & 0                                           & c_{\mathrm{u}} 
              \end{array} \right) , \nonumber \\
M_{\mathrm{d}} &=& \left( \begin{array}{ccc}
      a_{\mathrm{d}} & d_{\mathrm{d}} & 0         \\
      d_{\mathrm{d}} & b_{\mathrm{d}} & e_{\mathrm{d}} \\
      0         & e_{\mathrm{d}} & c_{\mathrm{d}} 
              \end{array} \right), 
\label{eqn:minimal_form}
\end{eqnarray}
where $a_{\mathrm{u}} \sim d_{\mathrm{u}}$, 
      $a_{\mathrm{d}} \sim e_{\mathrm{d}}$ 
and $\theta$ are real. 
These mass matrices lead 
(1,2) mixing for up part and (1,2) and (2,3) mixing for down part, 
so they are suitable to (\ref{eqn:KM_mass_ratios}). 
The above quark mass matrices are minimally generic form of the SM, 
that is, 
there are just 10 parameters and there is no unmeasurable parameter. 

Starting from the above form (\ref{eqn:minimal_form}), 
we do not lose any generalities, 
because arbitrary quark mass matrices can be transformed into the above form. 
Arbitrary up and down part quark mass matrices $\hat{M}_{\mathrm{u,d}}$ 
can be transformed into hermitian matrices 
$\hat{H}_{\mathrm{u,d}}$ 
without changing any physical contents. 
%
So if arbitrary hermitian quark mass matrices can be transformed 
into the above form (\ref{eqn:minimal_form}) 
without changing any physical contents, 
it means that any quark mass matrices can be transformed into 
the form (\ref{eqn:minimal_form}) 
and that the form does not lose any generalities. 
The transformation is the following: 
\begin{equation}
{\mathcal U}^{\dagger} \hat{H}_{\mathrm{u,d}} {\mathcal U} = M_{\mathrm{u,d}}, 
\end{equation}
where ${\mathcal U}$ is the unitary matrix 
which is determined in the following way. 
At first, ${\mathcal U}_{i3}$ ($i=1 \sim 3$) is determined 
to be one of eigenvectors of $\hat{H}_{\mathrm{u}}$. 
Then we can obtain 
\begin{equation}
M_{\mathrm{u ~ 13}}=M_{\mathrm{u ~ 23}}=
M_{\mathrm{u ~ 31}}=M_{\mathrm{u ~ 32}}=0
\end{equation}
with the help of the unitarity of ${\mathcal U}$. 
Next, ${\mathcal U}_{i1}$ is determined by satisfying the equations 
\begin{eqnarray}
{\mathcal U}_{i3}^{*} \hat{H}_{{\mathrm{d}} ~ ij} {\mathcal U}_{j1} &=& 0 , 
\label{eqn:U3HU1_0} \\
{\mathcal U}_{i3}^{*} {\mathcal U}_{i1} &=& 0 , 
\end{eqnarray}
where $i,j=1 \sim 3$. 
(\ref{eqn:U3HU1_0}) guarantees 
\begin{equation}
M_{\mathrm{d ~ 13}}=M_{\mathrm{d ~ 31}}=0.
\end{equation}
Then ${\mathcal U}_{i2}$ is determined automatically 
because of unitarity of ${\mathcal U}$. 
Since ${\mathcal U}$ is unitary matrix, 
quark masses do not depend on the choice of ${\mathcal U}$. 
Moreover, the measurements of the KM matrix do not change, 
because 
both up and down part are transformed by the same unitary matrix ${\mathcal U}$. 

The mass matrices (\ref{eqn:minimal_form}) are minimally generic one, 
so we can express 
$a_{\mathrm{u}} \sim d_{\mathrm{u}}$, 
$a_{\mathrm{d}} \sim e_{\mathrm{d}}$ 
and $\theta$ 
in terms of physical contents. 
We can choose $\hat{H}_{\mathrm{u,d}}$ as 
\begin{eqnarray}
\hat{H}_{\mathrm{u}} &=& \left( \begin{array}{ccc}
                           m_{\mathrm{u}1} & 0 & 0 \\
                           0 & m_{\mathrm{u}2} & 0 \\
                           0 & 0 & m_{\mathrm{u}3} 
                    \end{array}\right) \mbox , \nonumber \\
\hat{H}_{\mathrm{d}} &=& V_{\mathrm{KM}}
                    \left( \begin{array}{ccc}
                           m_{\mathrm{d}1} & 0 & 0 \\
                           0 & m_{\mathrm{d}2} & 0 \\
                           0 & 0 & m_{\mathrm{d}3} 
                    \end{array}\right) 
                    V_{\mathrm{KM}}^{\dagger} \mbox , 
\label{eqn:def_H}
\end{eqnarray}
where $V_{\mathrm{KM}}$ is the KM matrix 
and $|m_{\mathrm{u(d)}i}|$ ($i=1 \sim 3$) is the $i$-th generation 
up (down) part quark mass. 
These mass matrices produce quark masses and the KM matrix accurately. 
Starting from (\ref{eqn:def_H}) and using the Wolfenstein form \cite{Wolf} 
as the KM matrix: 
\begin{equation}
V_{\mathrm{Wolfenstein}} \; \simeq \; \left ( \begin{array}{ccc} 
   \displaystyle 1-\frac{1}{2} \lambda^2 
      & \lambda 
         & A\lambda^3(\rho-{\mathrm{i}}\eta) \\ \\
   - \lambda 
      & \displaystyle 1-\frac{1}{2}\lambda^2 
         & A\lambda^2 \\ \\
   A\lambda^3(1-\rho-{\mathrm{i}}\eta) 
      & -A\lambda^2 
         & 1 
\end{array} \right ) \; ,
\end{equation}
we can obtain 
\begin{equation}
{\mathcal{U}} \simeq \left( \begin{array}{ccc}
\exp ( {\mathrm{i}} \phi_{1} ) 
  & \lambda \sqrt{\rho^2+\eta^2} \exp ( - {\mathrm{i}} \phi_{3} ) 
      & 0 \\
- \lambda \sqrt{\rho^2+\eta^2} \exp \{ {\mathrm{i}}(\phi_{1}+\phi_{3}) \} 
  & 1 & 0 \\
0 & 0 & 1
                     \end{array} \right) , 
\end{equation}
where 
\begin{eqnarray}
\phi_{1} &=& \arg ( 1-\rho+{\mathrm{i}}\eta ) , \nonumber \\ 
\phi_{3} &=& \arg ( \rho+{\mathrm{i}}\eta ) . 
\end{eqnarray}
Then we can obtain the leading terms of 
$a_{\mathrm{u}} \sim d_{\mathrm{u}}$, $a_{\mathrm{d}} \sim e_{\mathrm{d}}$ and $\theta$: 
\begin{eqnarray}
a_{\mathrm{u}} &\simeq & m_{{\mathrm{u}}1}
          +m_{{\mathrm{u}}2}\lambda^2(\rho^2+\eta^2), \nonumber \\
b_{\mathrm{u}} &\simeq & m_{{\mathrm{u}}2}, \nonumber \\
c_{\mathrm{u}} &=      & m_{{\mathrm{u}}3}, \nonumber \\
d_{\mathrm{u}} &\simeq & m_{{\mathrm{u}}2}\lambda(\rho^2+\eta^2) , \nonumber \\
a_{\mathrm{d}} &\simeq & m_{{\mathrm{d}}1}
          +m_{{\mathrm{d}}2}\lambda^2|1-\rho-{\mathrm{i}}\eta|^2, \nonumber \\
b_{\mathrm{d}} &\simeq & m_{{\mathrm{d}}2}, \nonumber \\
c_{\mathrm{d}} &\simeq & m_{{\mathrm{d}}3}, \nonumber \\
d_{\mathrm{d}} &\simeq & m_{{\mathrm{d}}2}
                         \lambda |1-\rho-{\mathrm{i}}\eta|, \nonumber \\
e_{\mathrm{d}} &\simeq & m_{{\mathrm{d}}3}A\lambda^2, \nonumber \\
\theta &\simeq & \arg \left( -\rho+{\mathrm{i}}\eta \right) 
               - \arg \left( 1-\rho+{\mathrm{i}}\eta \right). 
\end{eqnarray}
If we take 
\begin{eqnarray}
m_{{\mathrm{u}}1}=-m_u, 
  & m_{{\mathrm{u}}2}=m_c, 
    & m_{{\mathrm{u}}3}=m_t, \nonumber \\
m_{{\mathrm{d}}1}=-m_d, 
  & m_{{\mathrm{d}}2}=m_s, 
    & m_{{\mathrm{d}}3}=m_b, 
\end{eqnarray}
where $m_{u}$, $m_{d}$, $m_{c}$, $m_{s}$ $m_{t}$ and $m_{b}$ are 
the up, down, charm, strange, top and bottom quark masses, respectively, 
and make a certain phase rotation, 
we can obtain the quark mass matrices $M_{\mathrm{u,d}}$ as 
\begin{eqnarray}
M_{\mathrm{u}} &\simeq & \left(
                    \begin{array}{ccc} 
                      -m_u+m_c\lambda^2(\rho^2+\eta^2) 
                        & m_c\lambda(-\rho+{\mathrm{i}}\eta) 
                            & 0 \\
                      m_c\lambda(-\rho-{\mathrm{i}}\eta) 
                        & m_c 
                            & 0 \\
                      0 & 0 & m_t 
                    \end{array}
                    \right) \mbox , \nonumber \\
M_{\mathrm{d}} &\simeq & \left(
                    \begin{array}{ccc} 
                      -m_d+m_s\lambda^2|1-\rho-{\mathrm{i}}\eta|^2 
                        & m_s\lambda(1-\rho+{\mathrm{i}}\eta) 
                            & 0 \\
                      m_s\lambda(1-\rho-{\mathrm{i}}\eta) 
                        & m_s 
                            & m_b A \lambda^2 \\
                      0 & m_b A \lambda^2
                            & m_b 
                    \end{array}
                    \right) \mbox . \nonumber \\
 & & 
\label{eqn:parametrized_by_Wm}
\end{eqnarray}

To see the above matrices (\ref{eqn:parametrized_by_Wm}) more simply, 
let us introduce 
three parameters related to the measurements of the KM matrix, 
$p_{1}$, $p_{2}$ and $p_{3}$: 
\begin{eqnarray}
p_{1} &=& \frac{m_c\lambda^2(\rho^2+\eta^2)}{m_u} ,             \nonumber \\
p_{2} &=& \frac{m_s\lambda^2|1-\rho-{\mathrm{i}}\eta|^2}{m_d} , \nonumber \\
p_{3} &=& \frac{m_b A \lambda^2}{m_s} .
\label{eqn:def_parameters}
\end{eqnarray}
Using these parameters and making a certain phase rotation, 
we can write the matrices (\ref{eqn:parametrized_by_Wm}) as 
\begin{eqnarray}
M_{\mathrm{u}} &\simeq &
\left( \begin{array}{ccc} 
(p_{1}-1)m_{u} & \sqrt{p_{1}m_{u}m_{c}} \exp ({\mathrm{i}}\theta) & 0 \\
\sqrt{p_{1}m_{u}m_{c}} \exp (-{\mathrm{i}}\theta) & m_{c}         & 0 \\
0                                                 & 0             & m_{t} 
\end{array} \right) , 
\nonumber \\
M_{\mathrm{d}} &\simeq &
\left( \begin{array}{ccc} 
(p_{2}-1)m_{d} & \sqrt{p_{2}m_{d}m_{s}} & 0          \\
\sqrt{p_{2}m_{d}m_{s}} & m_{s}          & p_{3}m_{s} \\
0                      & p_{3}m_{s}     & m_{b} 
\end{array} \right) , 
\label{eqn:parametrized_by_pm}
\end{eqnarray}
where 
\begin{equation}
\theta \simeq  \arg \left(  -\rho+{\mathrm{i}}\eta \right) 
             - \arg \left( 1-\rho+{\mathrm{i}}\eta \right) 
\end{equation}
and $p_{1} \sim p_{3}$ and $\theta$ characterize the KM matrix. 
At this time, 
we can obtain 
\begin{eqnarray}
|V_{us}| 
&\simeq& 
\left|   \sqrt{\frac{p_{2}m_d}{m_s}} 
       - \sqrt{\frac{p_{1}m_u}{m_c}} \exp({\mathrm{i}} \theta ) \right| , 
\nonumber \\
|V_{cb}|
&\simeq& 
\frac{p_{3}m_s}{m_b} , 
\nonumber \\
\left| \frac{V_{ub}}{V_{cb}} \right| 
&\simeq& 
\sqrt{\frac{p_{1}m_u}{m_c}} , 
\nonumber \\
\left| \frac{V_{td}}{V_{ts}} \right| 
&\simeq& 
\sqrt{\frac{p_{2}m_d}{m_s}} . 
\end{eqnarray}

\section{Discussion}
\label{sec:Discussion}
\par
\hspace*{\parindent}
Using current experimental data, 
we can find that three parameters $p_{1} \sim p_{3}$ are nearly equal to 1. 
If we take current experimental data \cite{Koide_mass,PDG}
\begin{eqnarray}
\frac{m_u}{m_c} &=& 4.26 \times 10^{-3} \times (1 \pm 0.196) , \nonumber \\
\frac{m_d}{m_s} &=& 4.97 \times 10^{-2} \times (1 \pm 0.045) , \nonumber \\
\frac{m_s}{m_b} &=& 3.35 \times 10^{-2} \times (1 \pm 0.144) , 
\end{eqnarray}
and 
\begin{eqnarray}
\lambda &=& 0.217 \sim 0.224 , \nonumber \\
\lambda \sqrt{\rho^2+\eta^2} &=& 0.06 \sim 0.10 , \nonumber \\
A\lambda^2 &=& 0.035 \sim 0.042 , 
\end{eqnarray}
then we can obtain 
\begin{eqnarray}
p_{1} &=& \frac{m_c\lambda^2(\rho^2+\eta^2)}{m_u} 
                            = 0.71 \sim 2.92 \mbox , \nonumber \\
p_{2} &=& \frac{m_s\lambda^2|1-\rho-{\mathrm{i}}\eta|^2}{m_d} 
                            = 0.71 \sim 1.27 \mbox , \nonumber \\
p_{3} &=& \frac{m_b A \lambda^2}{m_s} 
                            = 0.91 \sim 1.46 \mbox .
\end{eqnarray}

It is instructive to consider energy dependence of 
the above parameters $p_{1} \sim p_{3}$. 
The parameters $p_{1}$ and $p_{2}$ in (\ref{eqn:def_parameters}) 
are almost energy independent, 
because $\rho$, $\eta$, $\lambda$, $m_u/m_c$ and $m_d/m_s$ are 
almost energy independent. 
As for the parameter $p_{3}$, 
$A$ has energy dependence, 
but $m_s/m_b$ has almost the same energy dependence 
in the case that $\tan \beta$ 
has magnitude of $O(1)$ \cite{Babu}. 
Then the parameter $p_{3}$ is energy independent 
in the case of $\tan \beta \sim O(1)$. 
Hence the parameters which are estimated at the weak scale 
are meaningful enough. 

The parameters $p_{1} \sim p_{3}$ are nearly equal to $1$. 
It means that 
the relations between the KM matrix elements and the quark mass ratios are 
understood in terms of the elements of the quark mass matrices. 
It is very interesting, 
because we might be able to regard the simple relations among the elements 
as the relations among the original Yukawa couplings. 
When $p_{1}$, $p_{2}$ equal to $1$, 
we obtain 
\begin{equation}
M_{\mathrm{u} ~ 11} \simeq 0 , ~~~~~ M_{\mathrm{d} ~ 11} \simeq 0 
\end{equation}
and 
\begin{eqnarray}
|V_{us}| 
&\simeq& 
\left|   \sqrt{\frac{m_d}{m_s}} 
       - \sqrt{\frac{m_u}{m_c}} \exp({\mathrm{i}} \theta ) \right| , 
\nonumber \\
\left| \frac{V_{ub}}{V_{cb}} \right| 
&\simeq& 
\sqrt{\frac{m_u}{m_c}} , 
\nonumber \\
\left| \frac{V_{td}}{V_{ts}} \right| 
&\simeq& 
\sqrt{\frac{m_d}{m_s}} . 
\end{eqnarray}
And then we obtain 
\begin{equation}
\arg(\rho+{\mathrm{i}}\eta) \simeq \frac{\pi}{2} , 
\end{equation}
taking account of 
\begin{equation}
\sqrt{\frac{m_d}{m_s}} \simeq \lambda 
               \gg \sqrt{\frac{m_u}{m_c}} 
               \gg \left| \sqrt{\frac{m_d}{m_s}}-\lambda \right| . 
\end{equation}
This point is similar to the Fritzsch {\itshape Ansatz} case \cite{Fritzsch1}. 
Moreover, 
when $p_{3}$ equals to $1$, 
we obtain 
\begin{equation}
M_{\mathrm{d} ~ 22} \simeq M_{\mathrm{d} ~ 23}  
                        ~ (M_{\mathrm{d} ~ 32}) 
\end{equation}
and 
\begin{equation}
|V_{cb}| \simeq \frac{m_s}{m_b} . 
\end{equation}
If all the parameters $p_{1}$, $p_{2}$ and $p_{3}$ are equal to $1$, 
the mass matrices (\ref{eqn:parametrized_by_pm}) 
become exceedingly simple form: 
\begin{eqnarray}
M_{\mathrm{u}} &=& \left( \begin{array}{ccc}
 0  & d_{\mathrm{u}}\exp{\left( {\mathrm{i}}\theta \right)}              & 0 \\
 d_{\mathrm{u}}\exp{\left( -{\mathrm{i}}\theta \right)} & b_{\mathrm{u}} & 0 \\
 0  & 0                                                 & c_{\mathrm{u}} 
                                    \end{array} \right) , \nonumber \\
M_{\mathrm{d}} &=& \left( \begin{array}{ccc}
                0              & d_{\mathrm{d}} & 0              \\
                d_{\mathrm{d}} & b_{\mathrm{d}} & b_{\mathrm{d}} \\
                0              & b_{\mathrm{d}} & c_{\mathrm{d}} 
                                    \end{array} \right) . 
\label{eqn:simple_form}
\end{eqnarray}
They have only 7 parameters, 
that is 6 hierarchical magnitudes of elements and 1 phase. 
These form might offer us the origin of the quark mass matrices. 

The top quark mass is very large, 
so (2,2) element and (2,3) element ((3,2) element) 
of the up part mass matrices 
might also take the same value. 
In this case, we obtain $V_{cb}$ as 
\begin{equation}
|V_{cb}| \simeq \frac{m_s}{m_b} \pm \frac{m_c}{m_t} ~ . 
\end{equation}

Note that the quark mass matrices 
which Fritzsch {\itshape et al.} have assumed in Ref. \cite{NewFritzsch} 
have different situation in $V_{cb}$. 
They start from "democratic mass matrix", 
that is, rank 1 matrix, 
and add a cirtain small parameters to it, 
then the mass matrices lead to 
\begin{equation}
|V_{cb}|
 \simeq \frac{1}{\sqrt{2}}\left( \frac{m_s}{m_b} \pm \frac{m_c}{m_t} \right). 
\end{equation}
In this case, 
the parameter $p_{3}$ is not nearly equal to $1$ 
but nearly equal to $1/\sqrt{2}$ because $m_c/m_t \ll m_s/m_b$. 

P. Ramond {\itshape et al.} have suggested 
the similar structure mass matrices 
to the above form (\ref{eqn:simple_form}) in Ref. \cite{Ramond}. 
But they do not have mentioned the possibility 
that the (2,2) element and (2,3) (or (3,2)) element of down part mass matrices 
are almost the same magnitudes each other in Ref. \cite{Ramond}. 

To explain 
$M_{\mathrm{d} ~ 22} \simeq M_{\mathrm{d} ~ 23} 
                           (M_{\mathrm{d} ~ 32})$ 
is a problem, 
even if we explain zero elements by flavor symmetry. 
If we regard the origins of 
$M_{\mathrm{d} ~ 22}$ and 
$M_{\mathrm{d} ~ 23}$ ($M_{\mathrm{d} ~ 32}$) 
as the same one Higgs vacuum expectation value and Yukawa coupling, 
and if the origin of $M_{\mathrm{Simpler ~ d} ~ 33}$ is different from it, 
it is difficult to explain them by giving charges to fields. 
Namely, 
$d_{\mathrm{L} ~ 2}$ 
(down part left handed quark field of the 2nd family) 
and $d_{\mathrm{L} ~ 3}$ 
(down part left handed quark field of the 3rd family) 
must have the same charge 
to couple with the same Higgs. 
Similarly, 
$d_{\mathrm{R} ~ 2}$ 
(down part right handed quark field of the 2nd family) 
and $d_{\mathrm{R} ~ 3}$ 
(down part right handed quark field of the 3rd family) 
must do so. 
On the other hand, $d_{\mathrm{L} ~ 3}$ and $d_{\mathrm{R} ~ 3}$ 
couple with other Higgs and it lead large (3,3) element of 
the down part mass matrix. 
At this time, 
(2,2), (2,3) and (3,2) elements of the down part mass matrix must be 
as large as (3,3) element of the down part mass matrix 
because $d_{\mathrm{L} ~ 2}$ and $d_{\mathrm{L} ~ 3}$ have the same charge 
and so do $d_{\mathrm{R} ~ 2}$ and $d_{\mathrm{R} ~ 3}$. 
This is caused by the same magnitude elements in the same column or row. 

This situation does not change 
when we assume 
that quark mass matrices have only 6 hierarchical elements except phases 
and that $|V_{cb}| \simeq m_s/m_b$ mainly comes from 
the second and the third columns and rows of down part mass matrix. 
If we suppose there is no same order element 
in the same column and the same row, 
only $M_{\mathrm{d} ~ 33}$ must be the largest 
to obtain different order quark masses. 
Then we obtain $M_{\mathrm{d} ~ 33} \simeq m_{b}$ 
from trace of $M_{\mathrm{d}}M_{\mathrm{d}}^{\dagger}$. 
So we can write the (2,3) part of $M_{\mathrm{d}}$ as 
\begin{equation}
M \simeq \left(\begin{array}{cc}
                   x & y     \\
                   z & m_{b} 
         \end{array}\right) ~, 
\end{equation}
where $x, y, z \ll m_{b}$. 
(We assume $x$, $y$ and $z$ are real for simplicity.) 
At this time, 
\begin{equation}
M M^{\mathrm{T}} \simeq \left(\begin{array}{cc}
                                  x^{2}+y^{2} & ym_{b}    \\
                                  ym_{b}      & m_{b}^{2} 
                        \end{array}\right) 
\end{equation}
must be almost diagonalized by 
\begin{equation}
\left(\begin{array}{cc}
   1                                  & \displaystyle \frac{m_{s}}{m_{b}} \\
   \displaystyle -\frac{m_{s}}{m_{b}} & 1                                   
\end{array}\right)
\end{equation}
in order to mainly obtain $|V_{cb}| \simeq m_s/m_b$ 
from the (2,3) part of $M_{\mathrm{d}}$. 
Hence we obtain $y \simeq m_{s}$. 
Then determinant of $M$ is 
\begin{equation}
xm_{b}-zm_{s} \simeq m_{s}m_{b} ~ . 
\end{equation}
So we obtain $x \simeq m_{s}$, 
taking account of $z \ll m_{b}$. 
However, this contradicts the supposition 
that there is no same order elements in the same column and the same row. 
Therefore, we need the same order elements at the same column or row 
in (2,3) part of $M_{\mathrm{d}}$ 
when we assume that $|V_{cb}| \simeq m_s/m_b$ mainly comes from 
(2,3) part of $M_{\mathrm{d}}$ 
and that only 6 hierarchical elements of quark mass matrices except phases. 
This is the very reason 
that the Fritzsch {\itshape Ansatz} \cite{Fritzsch1} 
can not explain $|V_{cb}|$ 
and that the Branco-Silva-type \cite{Branco2} can explain $|V_{cb}|$. 
Anyway, 
in order to explain $p_{3} \simeq 1$, 
we need a mechanism or something. 

\section{Summary}
\label{sec:Summary}
\par
\hspace*{\parindent}
We have studied relations between the quark masses 
and the KM matrix elements 
on the minimally generic quark mass matrices. 
They are hermitian matrices and have 3 texture zeros. 
Furthermore, 
they have only 10 parameters, 
so we can express all their elements in terms of only physical contents. 
Then we have seen 
that current experimental data suggest some interesting simple relations 
among the elements in the minimally generic form. 
In order to clear this point, we have introduced three parameters 
which are almost energy independent for $\tan \beta \sim O(1)$. 
The relations among the elements might offer us 
the relations among the original Yukawa couplings, 
so it is very interesting. 
In addition, we have seen the possibility 
that the relations 
$|V_{us}|\sim \sqrt{m_d/m_s}$, $|V_{cb}|\sim m_s/m_b$ and 
$|V_{ub}/V_{cb}|\sim \sqrt{m_u/m_c}$ 
are explained by only 6 hierarchical elements of mass matrices and 1 phase. 

Moreover, 
it has pointed out 
that to explain the relation $|V_{cb}|\sim m_s/m_b$ 
has difficulty in flavor symmetry 
if we assume 
that the quark mass matrices have only 6 hierarchical elements except phases 
and that $|V_{cb}| \simeq m_s/m_b$ mainly comes from down part. 
This situation does not change in general. 
So, in order to explain $|V_{cb}| \simeq m_s/m_b$, 
we might have to introduce a mechanism or something. 

In the near future, the three parameters will be made out more accurately 
in detailed experiments, for example experiments in b-factory. 
They will offer us important information 
about the origin of the quark masses and the KM matrix. 

\section*{Acknowledgements}
\par
\hspace*{\parindent}
This work was supported in part by 
JSPS Research Fellowships for Young Scientists.

\end{document}